# Phosphorene Heterostructure Nanodevices for Ultrafast Energy Harvesting and Next Generation Electronics

Sydney L. Marler
Phillips Academy Andover

Phosphorene Heterostructure Nanodevices for Ultrafast Energy Harvesting and Next Generation Electronics

Sydney Marler
Phillips Academy Andover, Andover, MA, USA Mentor: Dr. Jiang Wei

Transition metal dichalcogenides (TMDCs) have recently shown much promise as thin layer semiconductors in application to transistor technology. TMDCs provide a unique vantage point for studying the properties of phosphorene, a highly efficient and tunable 'super-material.' Phosphorene-TMDC heterostructure nanodevices have remained largely unexplored due to the lack of air stability observed in phosphorene under ambient conditions. This study investigates a novel nanofabrication technique that effectively enhances the air stability and practical scalability of graphene-analogous phosphorene. A phosphorene-$WS_2$ heterostructure was first designed and synthesized via micromechanical exfoliation and dry transfer methods. A novel 'umbrella contact' was fabricated using electron beam lithography which extended the electrical anode over the phosphorene heterojunction region. SEM, Raman spectroscopy, gate transport, and photocurrent response techniques were employed to characterize the device physically and electronically. Devices fabricated with an extended anodic contact remained functional for extended periods of time, suggesting that the contact had prevented the phosphorene from degradation in ambient conditions. Preliminary results obtained via current response indicate a promising heterojunction with high speed transistor properties. This work suggests the unprecedented longevity and practical scalability of phosphorene utilizing a novel contact nanofabrication technique. Potential applications of the nanodevice include next generation electronic devices, photodetectors, energy storage devices, and photovoltaic cells.

Acknowledgements

First and foremost, I would like to thank the Quantum Devices Laboratory at Tulane University and the Nano-optics Research Center at Harvard University for allowing me to conduct research using their facilities. I would also like to thank Dr. Jiang Wei, Dr. Marko Loncar, Mr. Chunlei Yue, and Mr. Xue Liu for their advice and encouragement throughout the research process.

# 1 Introduction

Since the discovery and application of new and exciting materials such as graphene and carbon nanotubes, the design and scalability of two dimensional materials has attracted a considerable amount of attention within the scientific community. However, several major difficulties have been found in the utilization of graphene, a very promising material for applications ranging from energy storage to flexible electronics that include a lack of mobility and toxicity to consumers from inhalation. Graphene and similar allotropes of versatile carbon sheets provide advantages in applications to the consumer electronic industry for their high speed conductivity, flexibility, and thin structure. Ultrathin transition metal dichalcogenides (TMDCs) have been the subject of much attention recently due to their emergent optical and electrical properties. TMDCs are represented as $MX_2$, including the transition metal (M) and the chalcogen (X), and have unique properties due to the weak Van der Waals interactions between 'sandwiched' metal between two chalcogen layers. TMDCs have ideal large electronic band gaps that allow for increased mobility and control that has only recently been explored in semiconducting materials. Previous literature has reported promising photoluminescence properties in several transition metal dichalcogenides, hinting at their promise for photovoltaic and energy harvesting applications. Useful TMDCs, where M is W or Mo and X is S, Se, or Te, vary in their air stability and sizable band gaps in first principles literature. TMDCs display characteristic optoelectronic properties when micromechanically converted from bulk to thin layered form. This allows for the characterization of materials via Raman spectroscopy, used later in this study.

Van der Waals heterostructures are created by stacking layers of two dimensional materials. This layering allows for unique device structures through which electrical transport provides new insight into device functionality. The differing materials create band offsets which provides a method for manipulating the properties of two dimensional materials. A heterostructure typically consists of an n-type material and a p-type material. P-type materials contain electrons and n-type materials carry corresponding holes, which creates an electron gradient. This is useful for both device conductivity and photovoltaic applications. The layers between the p- and n-types are bonded by weak van der Waals forces. Heterostructures can be synthesized using both dry and wet transfer methods, but this study will make use exclusively of dry transfer methods.

TMDCs have attracted much attention from previous literature as a promising p-type material (Tong et al). $WS_2$ and $MoS_2$ have been studied extensively and display promising electrical characteristics because of their band gaps (1.1-2eV). Past research has made extensive use of TMDC layers synthesized via mechanical exfoliation for stacking thin layers (Woods et al). TMDC heterostrcutures are commonly connected to a metallic contact (typically gold) using electron beam lithography. While displaying promising electrical properties, these devices lack in practicality due to costly and time consuming synthesis methods and insufficient TMDC electrical performance.

Phosphorene (black phosphorus) is an immensely promising material with a repeating honeycomb lattice shape of phosphorus units. In recent years, the material has attracted much attention due to its graphene analogous properties. Phosphorene features high hole mobility, flexibility, and strength. Unlike graphene, phosphorene has a band gap of 2 eV and an effective on-off ratio of 10^6. (You et al) Transition metal dichalcogenide/phosphorene heterostructure have rarely been studied, mainly due to the instability of phosphorene. Phosphorene degrades within hours in an ambient environment and relies on careful and enclosed procedures to retain its electrical properties. Previous studies have reported partial degradation in as little as 2 hours (while maintaining electrical properties) and total degradation in 48 hours (losing characteristic electrical properties. For this study, WS2 and MoS2 were studied intensively because they demonstrated the most promising range of band gaps in previous work (You et al).

Overall, phosphene has rarely been studied in the context of a heterostructure. In the past, there have been many attempts made at reducing the volatility of phosphorene in an ambient environment. These attempts include blanketing in a thick dielectric material (such as Boron Nitride) the phosphorene once it is synthesized into a device, and have been largely unsuccessful. In the most successful attempt, the phosphorene device maintained its electrical properties for 7 days. (Kim et al) The sample eventually decayed in the air due to uneven protection on the sides, as the dielectric blanketed the entire device. The device remained electrically ineffective because the blanketing dielectric insulated the phosphorene's contact points. While previous work somewhat improved air stability of phosphorene, practical and long-term applications in a functional semi conductive transistor remain a challenge.

This study approaches phoshphorene air-instability through the lens of heterostructure and contact synthesis. Phosphorene is a relatively novel material and little has been explored in the realms of TMDC-phsophorene heterostructure electrical properties. Much focus has been on making phosphorene viably and practically air-stable, so relatively little else has been explored. By constructing a device with phosphorene and WS2, chosen for its promising electrical properties and band gap compatibility with phosphorene, it becomes possible to use nano fabrication techniques to aid phosphorene in air stability. By synthesizing a heterostructure with dry transfer methods, it becomes possible to cover exclusively phosphorene without interfering with the electrical capabilities of the overall device (Figure1). This study focuses on the nano fabrication of an extended anodic contact over n-type phosphorene, leaving the p-type TMDC electrically functional. This novel approach allows for the indefinite air-stability of phosphorene in a practical and scalable nanodevice. Because no dielectric is required in this solution, the device additionally is functional as a photosensitive detector with promising applications in photovoltaic devices.

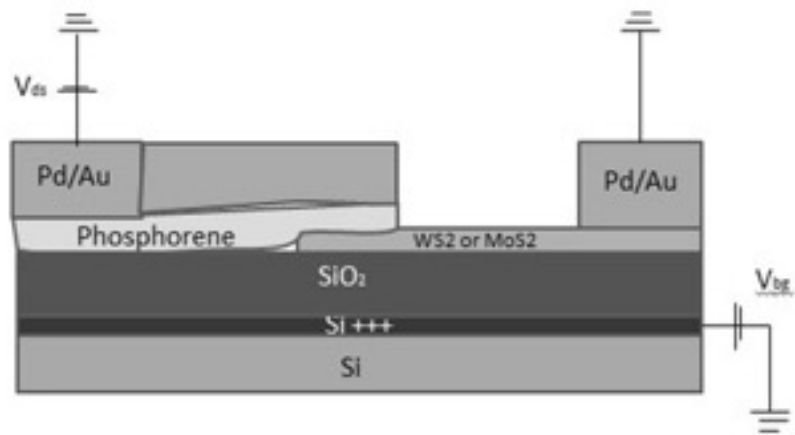

**Figure 1.** Device schematic of as fabricated FET. An extended anodic contact (Pd or Au) covers the air volatile phosphorene and heterostructure layering allows the device to retain functionality as a photodetector.

In order to increase the scalability in a cost-effective manner, palladium (Pd) contacts were integrated into the design of the Heterostructure device. This material decreases manufacturing costs by approximately 49% when compared to devices that utilize gold contacts. The Pd was shown in previous first-principles literature to be an effective contact during interactions with black phosphorus, effectively lowering the resistivity and decreasing the effects of the Schottsky barrier on the device's overall gate response performance.

Bulk quantity phosphorene has immense potential for scalability because of its advantageous crystalline structure that allows for simple and productive two-dimensional micromechanical exfoliation. Effective exfoliation methodology is a key step to scalable cost reduction in semiconductor nanodevices.

While TMDC materials have recently attracted much attention in the scientific community, much more work must be done to synthesize scalable semiconductor nano devices. Phosphorene based nano devices demonstrated the properties of a promising next-generation material. The device has applications in electronic and optoelectronic devices such as batteries, chemical sensors, personal electronic devices, photodetectors, and photovoltaic devices. In the scientific community, the implications of a phosphorene device extends to spintronics and quantum mechanics.

2  Methods

The nanostructure in this study were fabricated using a "top-down" approach in which a bulk crystal is micro mechanically exfoliated into a few layer thin film. Synthesis of bulk crystal tungsten disulfide, molybdenum disulfide, and black phosphorus (phosphorene) was performed using chemical vapor deposition (CVD). The general process primarily used inert iodine as a carrier for deposition, with variations to the process depending on the material being synthesized. Under a high temperature environment of 750 degrees celsius, the inert argon gas speeds up material growth by transferring kinetic energy to the desired material. The argon precursor spontaneously decomposes upon contact with the material. Semiconductor materials, such as WS2 and MoS2, were synthesized using atomic layer deposition (ALD). Using iodine as the initial precursor, a tradition metal layer was first synthesized. The surrounding chalcogen layers that sandwich the transition metal layer were synthesized using an argon precursor following the purge of iodine in the CVD chamber. Silicon dioxide platforms were created using plasma etched chemical vapor deposition (PECVD). PECVD utilizes ionized plasma to accelerate the reaction of the iodine precursor with the bulk SiO2, allowing for the creation of a heavily ionized silicon layer to be used in the gating process. Tungsten disulfide and molybdenum disulfide bulk crystal was synthesized into thin-layer form using micromechanical exfoliation processes. The bulk material was converted into a thin layer sheet by utilizing sticky tape to pull apart bulk layers. Black phosphorene bulk crystal was stored using a protective acetone solution prior to fabrication to prevent air degradation.

  The first step of fabrication involved the shaping of a silicon dioxide platform. A sheet of multilayer silicon-based material (heavily ionized silicon, elemental neutral silicon, and silicon dioxide, from bottom to top) was cut into a 1 cm by 1 cm shape. The finished wafer was then cleaned using an acetone bath, removing organic residue.

Micro mechanically exfoliated thin layer WS2 was transferred onto the wafer. From the exfoliated WS2 thin flakes few layer flakes were identified using color microscopy and chosen for fabrication.

In a protected environment to prevent ambient air degradation, phosphene (black phosphorus) was exfoliated onto sticky PDMS. Positioned over the silicon wafer with exfoliated WS2 flakes using a nanoscale positioning device, thin layer phosphorene flakes were identified and carefully lowered onto few layer WS2 until the dry transfer was complete. The success of the dry transfer was confirmed using an optical microscope. Next, the wafer containing the device was primed for electron beam lithography using methylisobutylketone (MIBK) and polymethylmethacrylate (PMMA). PMMA served to protect the nanostructure from damage and ambient air degradation, while the MIBK was employed to function as a chemical layer that would be removed by the electron beam lithography.

Scanning electron microscopy (SEM) was used analogously with electron beam (E-Beam) lithography in order to design and write a pattern for the contacts of the device. Contacts were designed using images from color microscopy to cover the few layer phosphorene in its entirety. SEM was used to view the surface of the device and to calibrate the instrument. The E-beam lithography took electrons from source, energized through a condenser in order to remove the MIBK layer from the device. This would later allow the electron beam evaporator to fabricate metallic contacts around the nanostructure.

E-beam evaporation allowed the deposition of Pd on to the device, in areas where the MIBK was decomposed by electron beam lithography. A crucible of bulk Pd was placed at the center of a chamber and melted. Accelerated elections were guided using a beam to the target.

The MIBK protective film and excess palladium were stripped from the device using acetone cleansing. Next, bonding of wires to the contacts and metallic strips allowed for the testing of the device. A back gate "scratch" in the multilayered silicon wafer was connected to the device to.

The completed device was then connected to a semiconductor analyzer, where charge was conducted through the device. The data confirmed the functionality of the phosphorene-based device. It also collected time-dependent conductance curves, indicating data regarding the efficiency and speed of the device. Raman spectroscopy was utilized to characterize the WS2/phosphorene heterojunction. Data was also collected using AFM, crystation, SEM, and optical microscope.

## 3 Results

To characterize the electrical properties of the device, a three terminal configuration was used with heavily doped silicon as the back gate electrode. The data indicated that the heterojunction was successful in maintaining electrical switching behaviors. Figure 3 demonstrates a measurements of source drain current (Ids) dependent on back gate voltage (Vbg) under ambient conditions. Vbg was set to a range of -1.6 V to 1.6 V.  The device has a characteristic curve of heterostructure devices in which the graph splits into separate electrical curves as electrons flow from the p-type to n-type. The on-off ration of the device reaches $10^6$, which suggests that the heterostructure has exceptionally high mobility when compared to graphene, WS2, and phosphorene exclusive devices.

**Figure 1.** Gate response configuration yielded a heterostructure characteristic curve over a -1.6 V to 1.6 V interval.

Voltage gate induced carrier density can be calculated using $n_{2D} = C_{ox}(V_{bg} - V_{bg,th})/e$ where $C_{ox}$ is the dielectric capacitance per unit area and $V_{bg,th}$ is the threshold voltage and $e$ is the unit charge. $C_{ox}$. The device displayed an approximate carrier density of 3.1

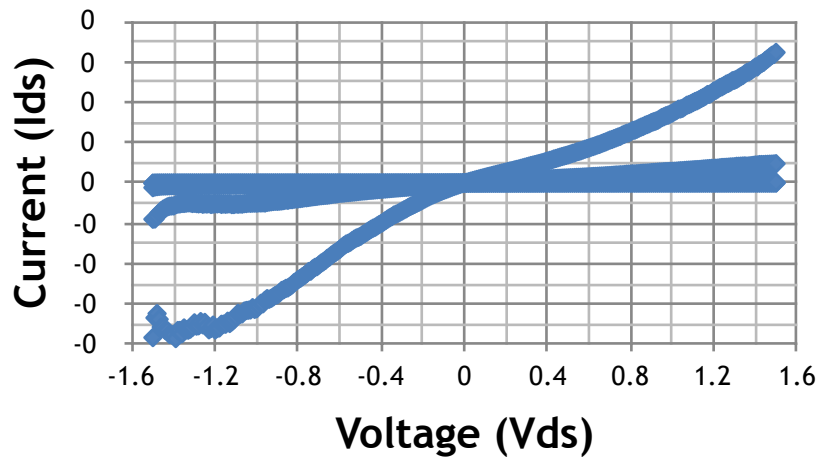

x $10^{12}$ $cm^{-2}$ at the displayed voltage range.

In order to study the transport mechanisms of the device, low temperature measurements were performed using a cryostat ion. The back gate voltage was fixed at 60 V and Ids- V ds sweeps were performed (Figure 4). This revealed the overall speed and efficiency of gate transport mechanisms. Temperature dependence was caused due to to the existence of the Schottky barrier. Only a small and statistically insignificant temperature dependence was found, indicating that the device has a highly functional on-off ratio.

In order to observe the control rate of degradation in a phosphorene sample left in an ambient enviornment, a thin flake of phsophorene was observed over the course of 144 hours. Degradation in the sample (Figure 2) was physically evident after just 24 hours and electrically evident after 48. The physical decay process involved the formation of oxygen bubbles, particularly around the edges of the device. An optical image shows that relatively thinner flakes (those used for nano fabrication) decay more quickly than their thicker counterparts.

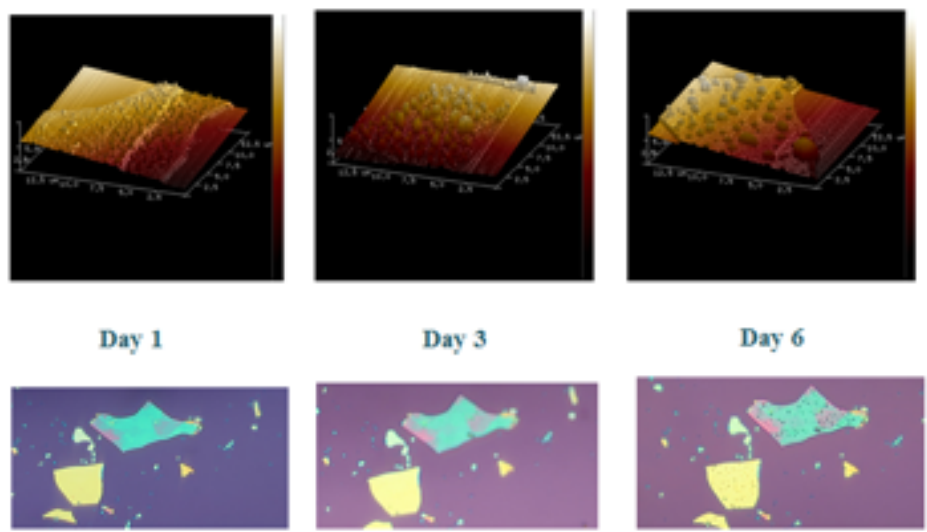

**Figure 2.** Top: AFM demonstration of phosphorene decay during a period of 144 hours in an ambient environment. Bottom: Optical images of a phosphorene thin flake (blue) used for AFM analysis

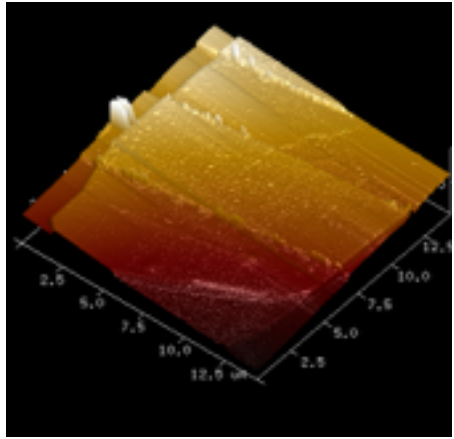

**Figure 3.** Atomic force microscopy analysis of phosphorene underneath and extended anodic palladium contact.

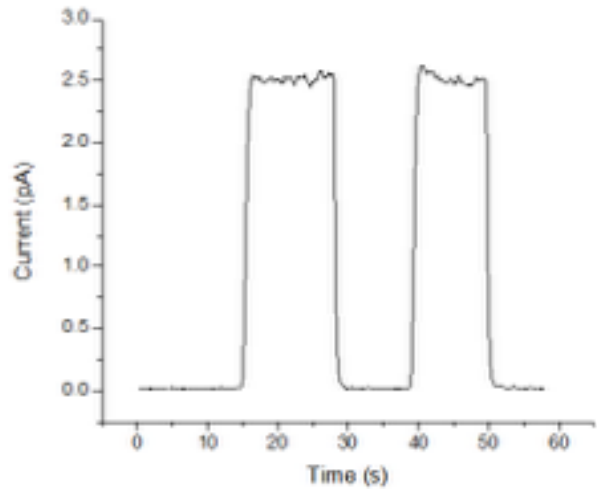

**Figure 4.** Photocurrent response of device. A light was switched on and off over an interval. The response can be measured in pico amps.

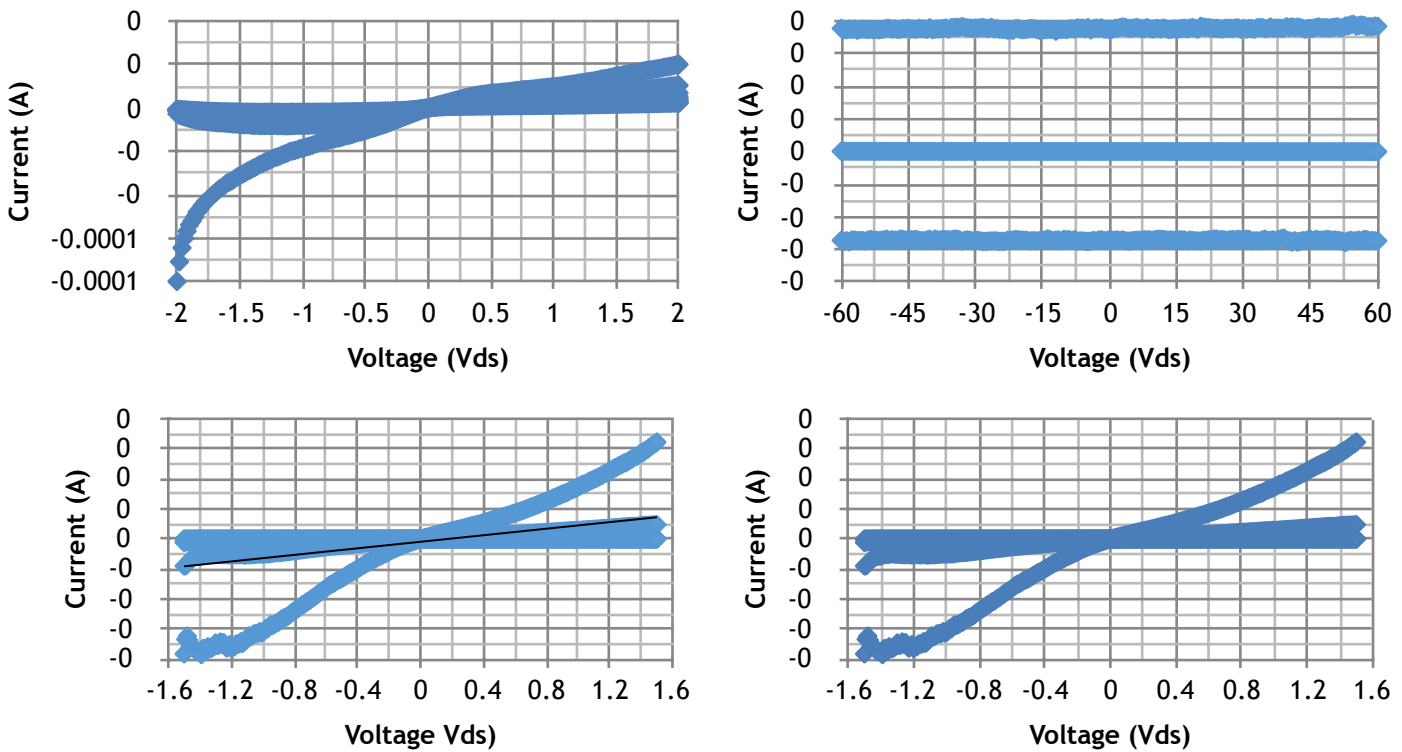

**Figure 5. a)** (Top) A phosphorene nano device with no extended anodic contact was used for the control group. The data points were collected 336 hours apart and all samples were stored in an ambient environment. **b)** (Bottom) A heterostructure with phosphorene and an anodic contact. The heterostructure remains intact after the same period of time.

Because the device is photovoltaic, photocurrent measurements taken with an Ids-Vds sweep reveals the speed and efficiency of the heterojunction as as function of time.

A control device was synthesized using only phosphorene as the semiconductor material and allowed to sit in an ambient environment for 336 hours. This device was then compared to the phosphorene heterostructure device with the extended palladium anodic contact under the same conditions and time. A three channel back gate response method was utilized to compare electrical characterizations. The phosphorene device was found to have lost all electrical properties due to degradation. The extended contact device was able to closely maintain electrical characteristics. This suggests that the phosphorene in the heterostructure had not yet degraded in the ambient environment. While it is unknown how indefinite this preservation of phosphorene is, the 336 hours currently surpasses previous works' air stability of phosphorene by 700%.

Raman spectroscopy was employed in characterizing the heterostructure device. Spectral lines for both WS2 and phosphorene can be clearly identified in the spectral signature. This is useful for characterizing the novel material for future studies.

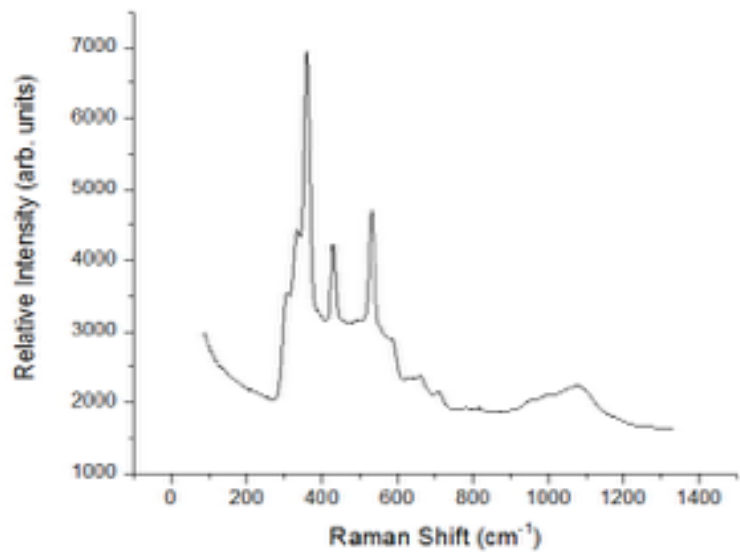

**Figure 6.** Spectral signatures for WS2 and Phosphorene

# 4 Conclusion

This preliminary study explores and characterizes the van der Waals heterojunction characterization of phosphorene/TMDC devices and proposes a novel nanofabrication technique to create electrically effective air stable thin layer phosphorene devices for practical applications. This research demonstrates the simple synthesis of heterostructure devices using micromechanical exfoliation and dry transfer methods.

Both TMDC synthesized heterojunction devices demonstrate successful photovoltaic and energy harvesting properties in ambient, vacuum, and temperature dependent settings. Raman characterization clearly shows peaks of all present materials at the heterojunction. Electrical response displays excellent mobility for both unique heterostructure devices.

A practical way to produce air stable phosphorene holds much promise as a method to potentially utilize a material exceeding analogous graphene in applications ranging from photovoltaic devices to flexible electronics. This work introduces a novel idea to address this challenge and future studies regarding phosphorene should be pursued.

Further work should explore more detailed characterization of phosphorene heterostructures using Raman spectroscopy and work to utilize interesting photovoltaic and photo-sensing properties of heterostructure phosphorene devices that was found in the study. Scalable techniques may assist in applicability of phosphorene based nanodevices.

The worldwide energy crisis is one of the most urgent and wide reaching problems that we as a society must face today. There has never been more demand for increased speed and efficiency in energy harvesting devices, particularly in renewable energy collection mechanisms. Integrating efficient and innovative devices into electronic systems maximizes the usage of energy and allows for top speed and efficiency in energy harvesting. The nanodevice uses a novel concept in nanofabrication called the 'umbrella contact' that provides a shield for phosphorene in heterostructure nanodevice settings that makes it among the first of its kind to be air stable in a practical setting.

The nanodevice shows immense potential for phosphorene to be integrated into a broad field of applications including electronic devices, wearable technology, biomedical sensors, and photovoltaic solar cells. Phosphorene based 'umbrella contact' nanoscale devices are simple to create, low cost, and show high potential for scalability and manufacturing. As a unique fabrication technique, the utilization of air stable phosphorene in nanodevices shows promise to benefit society by making a crucial step forward in energy management and the development of the next generation of electronic devices.